\documentclass{ulg_ptf}

\usepackage{epsfig}

\usepackage{amssymb}
\usepackage{amsmath}
\usepackage{float}
\usepackage{subfigure}
\newcommand{\vect}[1]{\vec{#1}}

\newcommand{\eqs}[1]{\begin{eqnarray}#1\end{eqnarray} }

\newcommand{\matrice}[1]{\begin{matrix} #1\end{matrix} }
\newcommand{\ks}[1]{#1 \!\!\! \slash } 

\newcommand{\nn}{\nonumber}
\newcommand{\cal}{\mathcal}

\begin{document}

\begin{frontmatter}





\title{A Model for the Off-forward Structure Functions of the Pion}


\author[a,b,*]{F. Bissey}, \author[a]{J. R. Cudell}, \author[a]{J. Cugnon},
\author[a,IISN]{J. P. Lansberg} and \author[a,FNRS]{P. Stassart}

\address[a]{D\'epartement de  Physique, B\^at. B5a, Universit\'e de  Li\`ege, 
Sart Tilman, B-4000 Li\`ege 1, Belgium}
\address[b]{Institute of Fundamental Sciences, Massey University, Private Bag 11 222, 
Palmerston North, New Zealand}

\thanks[*]{E-mail addresses: F.R.Bissey@massey.ac.nz, JR.Cudell @ulg.ac.be,
J.Cugnon@ulg.ac.be, JPH.Lansberg@ulg.ac.be, Pierre.Stassart@ulg.ac.be}
\thanks[IISN]{IISN Research Fellow}
\thanks[FNRS]{FNRS Research Associate}
\begin{abstract}
We extend our model for the pion, which we used previously to calculate its diagonal 
structure function, to the off-forward case.
The imaginary part of the off-forward $\gamma^{\star} \pi \rightarrow
\gamma^{\star}\pi$ scattering amplitude is evaluated in the chiral limit ($m_\pi=0$) 
and related to the twist-two and twist-three
generalised parton distributions ${H}$, ${H}^3$, $\tilde {H}^3$. 
Non-perturbative effects, linked to the size of the pion and still preserving 
gauge invariance, are included. 
Remarkable new relations 
between ${H}$, ${H}^3$ and $\tilde {H}^3$
are obtained and discussed.

\end{abstract}

\end{frontmatter}

% main text

 \section{Introduction}

Structure functions, which can be extracted from deep-inelastic 
experiments, are useful tools to understand the structure of
hadrons. Even if their $Q^2$ evolution is consistent with perturbative QCD, 
they result mainly from non-perturbative effects that are still not calculable
in the framework of QCD. This has led to phenomenological quark models
embodying various non-perturbative aspects of QCD. These models can be used to depict
 the behaviour of the structure functions and to understand the
connection between data and non-perturbative aspects of hadrons.
There has been extensive work on diagonal distributions along these lines 
(see Refs~\cite{ours}-\cite{diagonal} for the pion case). These distributions
can be used as the initial condition for a DGLAP evolution, which is necessary 
before a comparison with data~\cite{experim}. Such models can be applied
to the off-diagonal case, for which 
generalised structure functions~\cite{extramuller} can be linked~\cite{Belitsky:2000vk}
to generalised parton distributions
(GPD's).

One of the setbacks of phenomenological quark models suited to the description of 
the low-energy features of hadrons is that the underlying quark structure 
is obscured by the necessary introduction of regularisation procedures which
result in non-negligible differences in the structure functions.
 
To avoid these complications, we investigated the diagonal structure functions
in the case of a simple model for the pion~\cite{ours}, where the 
pion-quark-antiquark pseudoscalar coupling ($i\gamma^5 g_{\pi q\bar q}$) 
yields the correct symmetry, 
while the non-perturbative aspects come from a momentum cut-off mimicking 
the size of the pion, but still preserving gauge invariance.
This freed us from the question of what would be the detailed inner structure
of the meson. In that calculation, owing to the introduction of such a cut-off, 
crossed diagrams for the pion-photon scattering appear as higher
twists, leading twist structure functions can be identified,
and a reduction of the momentum fraction carried by the quarks is observed. Of course, 
as the cut-off is relaxed to let the quarks
behave freely, the momentum sum rule $\left<2x\right>=1$ is recovered at infinite $Q^2$.
Having that tool at hand, we now turn to  the
investigation of the properties of off-diagonal parton distributions, which are
likely to  shed some light on parton correlations and which have therefore
attracted much interest in recent years~\cite{JI97,RA96,CO97,GPD_review,GPD_pion1,GPD_pion}.

In the following, we  calculate the imaginary part of the
off-forward photon-pion scattering amplitude,
and of the structure functions $F_1,\dots,\ F_5$, related to 
the five independent
tensor structures in the scattering amplitude, and we
discuss their behaviour. We relate them to vector and axial vector form factors
and to the twist-two 
and twist-three generalised parton distributions (GPD's)  
${H}$, ${H}^3$ and $\tilde {H}^3$. We shall show that, within 
our model and in the high-$Q^2$ limit, the non-diagonal structure functions 
$F_3$ and $F_4$ are related to $F_1$, while $F_5$ happens to be a higher twist.
These results lead to new relations for the GPD's in the neutral pion case.

\section{General tensorial structure of the  $\gamma^{\star} \pi \rightarrow 
\gamma^{\star}\pi$ amplitude }
\subsection{External Kinematics}\label{subsec:kinematics}

\begin{figure*}
\centering
\mbox{
      \subfigure[]{\psfig{figure=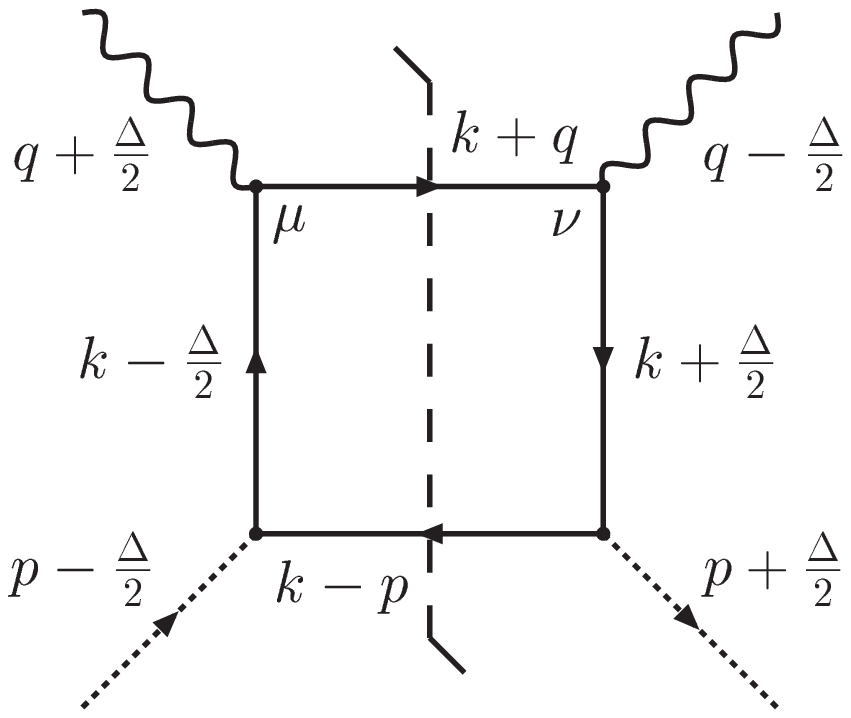,height=2.5cm}}\quad
      \subfigure[]{\psfig{figure=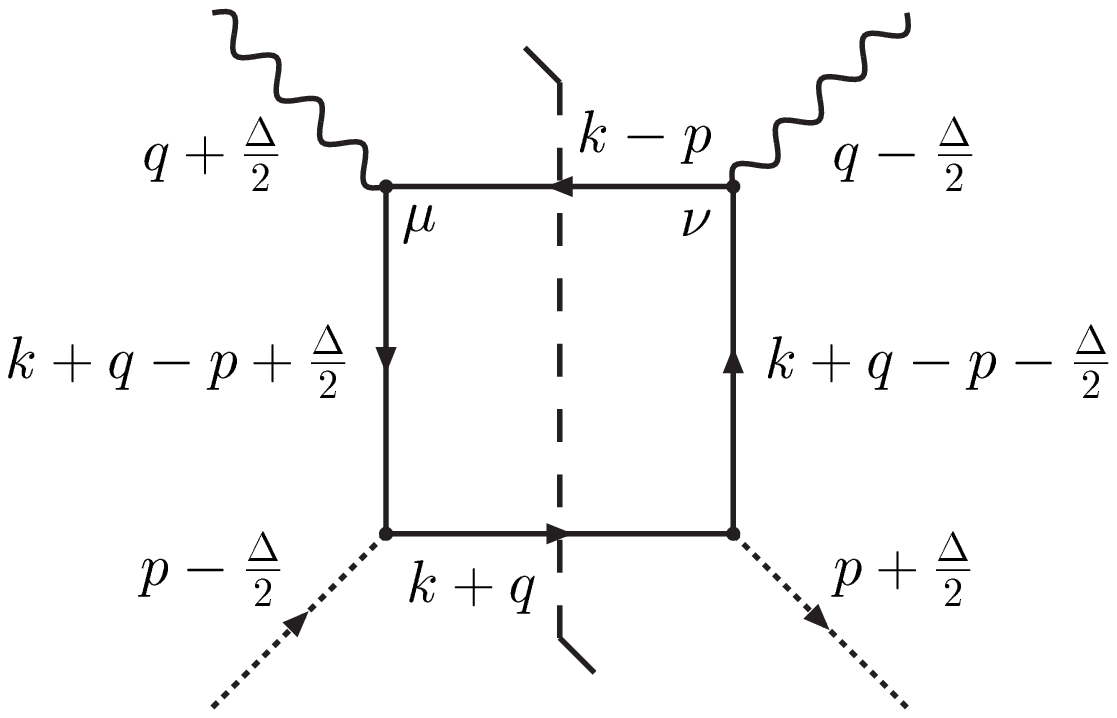,height=2.5cm}}}
\mbox{\subfigure[]{\psfig{figure=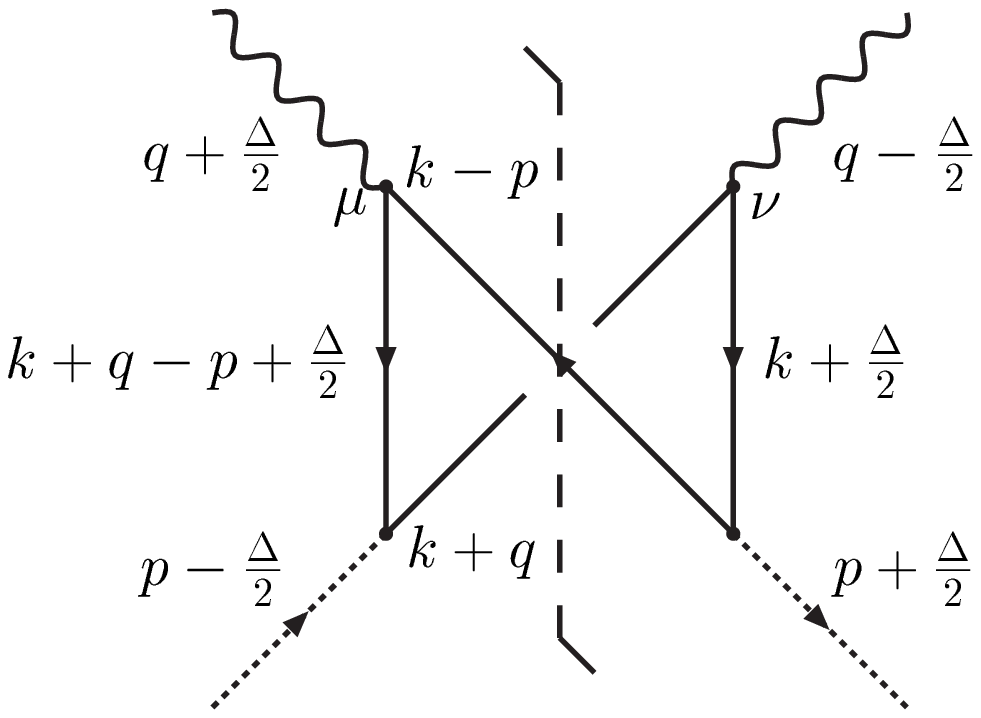,height=2.5cm}}\quad\quad\quad
      \subfigure[]{\psfig{figure=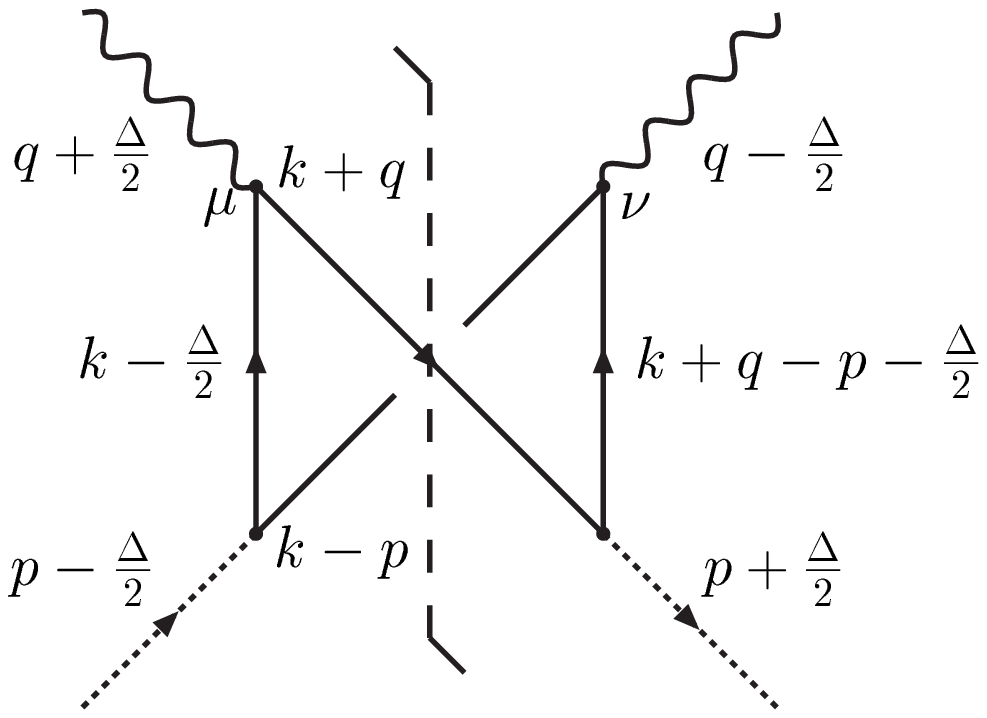,height=2.5cm}}}
\caption{The simplest diagrams contributing to the imaginary part of the 
amplitude for the scattering $\gamma^{\star} \pi \rightarrow 
\gamma^{\star}\pi$. Upper
(lower) diagrams are referred to  as box (crossed) diagrams. Dashed lines
represent the discontinuity of the amplitudes, {\it i.e.} their imaginary parts.}
\label{fig:diagrams}
\end{figure*}

Let $p_1$ and $p_2$ be the momenta of the ingoing and outgoing pions, $q_1$ and $q_2$ 
those of the corresponding photons (see Fig.~\ref{fig:diagrams}). 
Defining $p = \frac{(p_1 + p_2)}{2}$, $q = \frac{(q_1 + q_2)}{2}$ and $\Delta = p_2 -
p_1$, one can then write the scattering amplitude as a function of 
the Lorentz invariants $t = \Delta^2$,  $Q^2 = -q^2$, 
$x = \frac{Q^2}{2p\cdot q}$ and $\xi = \frac{\Delta\cdot q} {2p\cdot q}$.

In the elastic limit, characterised by $(q+\frac{\Delta}{2})^2=(q-\frac{\Delta}{2})^2$, 
one has $\Delta\cdot q = 0$ 
and thus $\xi=0$, while the diagonal limit ($\Delta = 0$) is obtained for $\xi=0$ and $t=0$. We further
recover the Bjorken variable $x = x_B$, where $ x_B = -\frac{q_1^2}{2p_1\cdot q_1}$. 

For Virtual Compton Scattering (VCS), for which the outgoing photon is on-shell, $\xi$ is 
related to $x$ through $\xi = -x \left(1 - \frac{\Delta^2}{4 Q^2}\right)$. 
Hence in the Deeply Virtual Compton Scattering (DVCS) limit, $t \ll Q^2$ and $\xi = -x$. 

\subsection{The structure functions $F_i$'s}
The hadronic tensor is defined through
\begin{eqnarray}
\label{deftmunu}
T_{\mu\nu} (q, p, \Delta) =
i \!\int \!d^4r {\rm e}^{i r \cdot q}
\left< p_2 | T j_\mu (r/2) j_\nu (-r/2) | p_1 \right>.
\end{eqnarray}

There exist five independent kinematical structures in Eq.\ (\ref{deftmunu}) 
that parametrise the photon-pion amplitude. Defining the projector
${\cal P}_{\mu\nu} = g_{\mu\nu} - \frac{q_{2 \mu} q_{1 \nu}} {q_1 \cdot q_2}$ and making use of these
five structures, we can rewrite $T_{\mu\nu}$ as follows:
\begin{eqnarray}
\label{tmunudecomp}
T_{\mu\nu}(q, p, \Delta) =
- {\cal P}_{\mu\sigma} g^{\sigma\tau} {\cal P}_{\tau\nu}
F_1
+ \frac{{\cal P}_{\mu\sigma} p^\sigma p^\tau {\cal P}_{\tau\nu}}{p \cdot q}
F_2\nonumber\\
+ \frac{{\cal P}_{\mu\sigma} (p^\sigma (\Delta^\tau-2\xi p^\tau)
+  (\Delta^\sigma-2\xi p^\sigma) p^\tau) {\cal P}_{\tau\nu}}{2 p \cdot q}
F_3\nonumber\\
+ \frac{{\cal P}_{\mu\sigma}
(p^\sigma (\Delta^\tau-2\xi p^\tau)
-  (\Delta^\sigma-2\xi p^\sigma)p^\tau) {\cal P}_{\tau\nu}}{2 p \cdot q}
F_4\nonumber\\
+ {\cal P}_{\mu\sigma}
(\Delta^\sigma-2\xi p^\sigma) (\Delta^\tau-2\xi p^\tau) {\cal P}_{\tau\nu}
F_5    .
\end{eqnarray}

Current conservation is ensured by means of the projector 
${\cal P}_{\mu\nu}$. Our notation slightly differs from Ref.~\cite{Belitsky:2000vk}: 
we have included a factor $1/m_\pi^2$ in the definition 
of $F_5$ in order to avoid divergences when the chiral limit is taken.
Note that Bose symmetry requires $F_1$, $F_2$,
$F_4$, $F_5$ to be even and $F_3$ to be odd in $\xi$.

\section{The model}

\subsection{General description}

We use the pion model introduced in our previous work~\cite{ours},
in which the $q \overline{q} \pi$ vertex is represented by the simplest
pseudoscalar coupling. The Lagrangian includes massive pion and massive quark
fields interacting through the pseudoscalar vertex, with an effective
pion-quark coupling constant.

Considering an isospin
triplet pion field $\vec{\pi}=(\pi^+,\pi^0,\pi^-)$ interacting with quark
fields $\psi$ the  Lagrangian density reads
\eqs{
 {\mathcal L}_{int}= i g (\overline{\psi}\mbox{  $\vec{\tau}$} \gamma_5 \psi) .
\mbox{  $\vec{\pi}$}, \label{Lint}
}
where $\vec{\tau}$ is the isospin vector operator. 

Of course, if our pseudoscalar field is to represent real pions, we have to 
impose that the corresponding hadrons have a finite size. That we shall do through
the use of a cut-off, as detailed below, the choice of which sets a
constraint on the value of the quark-pion coupling constant~\cite{ours}.

We shall limit ourselves in this paper to the calculation of the imaginary part
of the scattering amplitude, which allows a direct comparison with our previous
work and which is sufficient to determine the GPD's of neutral pions \cite{Belitsky:2000vk}.

At the leading order in the loop expansion, four diagrams contribute. They 
are displayed in Fig.~\ref{fig:diagrams}. Following the kinematics defined in 
section~\ref{subsec:kinematics} and applying Feynman rules, 
it is straightforward to write down the analytical expression for the scattering amplitude.
For a given set $\mu,\nu$ of the photon indices and with well-known 
conventions\footnote{The isospin/charge factor $(e_u^2+e_d^2)$ corresponds to the following 
choice of the isospin matrix: $\pi^-:\left(\matrice{ 0 & 0 \\ \sqrt{2} &0}\right)$ ; 
$\pi^0:\left(\matrice{ 1 & 0\\ 0 &-1}\right)$ ; $\pi^+:\left(\matrice{ 0 &   \sqrt{2}\\ 0 &0}\right)$ ; 
$\gamma:\left(\matrice{ e_u&  0 \\ 0&e_d}\right)$  }, the contribution of the first diagram
(a) shown in Fig.~\ref{fig:diagrams} to the scattering amplitude reads

\eqs{\label{eq:ampl_gpiqq}
{\cal M}^{\mu\nu}_a= 
3 g^2 (e_u^2+e_d^2)\int {\rm d}^4k 
{\rm Tr} (\gamma^5  ((\ks k-\ks p)+m_q)\gamma^5 \nn\\
\frac{\ks k+\frac{\ks \Delta}{2}+m_q}{(k+\frac{\Delta}{2})^2-m_q^2}\gamma^{\nu} 
((\ks k+\ks q)+m_q) \gamma^\mu \frac{\ks k-\frac{\ks \Delta}{2}+m_q}
{(k-\frac{\Delta}{2})^2-m_q^2}).
}

Expressions from the other three diagrams of Fig.~\ref{fig:diagrams} --
the one with reverse loop-momentum and the two crossed diagrams -- are similar and are 
not written down. Results below all pertain to the chiral limit $m_\pi = 0$.

\subsection{The implementation of the cut-off}

A simple way to impose that the pion has a finite size  
is to require that the square of the relative four-momentum of the quarks
inside the pion is limited to a maximum value $\Lambda^2$. 
Before writing this explicitly, let us give the details of the 
internal kinematics, {\it i.e.} the one involving the loop-momentum $k$. 
Let $\phi$ and $\theta$ be the spherical angles of $\vect k$ with respect
to the $z$-axis taken as the direction of the incoming photon. Defining 
$k_\rho^2=|\vect k|^2$ and $\tau=k^2=k^2_0-k_\rho^2$, and using 
spherical coordinates, we write the element  of integration as:
\eqs{d^4k=dk_0 dk_\rho k_\rho^2 d(\cos \theta) d\phi,}
or with the help of the variable $\tau$:
\eqs{d^4k=dk_0 \frac{k_\rho}{2} d\tau d(\cos \theta) d\phi.}

According to Cutkosky rules, the imaginary part of the amplitude is 
obtained by putting the intermediate quark lines on shell. This is realised by
the introduction of the two delta functions,
$\delta\left((k+q)^2-m_q^2\right)$ and $\delta\left((k-p)^2-m_q^2\right)$.

Working out the delta functions, we obtain that:
\eqs{&\delta\left((k+q)^2-m_q^2\right)\delta\left((k-p)^2-m_q^2\right)=\nn\\
&\frac{1}{2k_\rho |\vect q|}\delta\left(\cos\theta-\cos\theta_0\right)
\frac{1}{2\sqrt{s}}\delta\left(k_0-k'_0\right)
}
with $\cos\theta_0=\frac{2k_0q_0-Q^2-m_q^2+\tau}{2k_\rho |\vect q|}$ and 
$k'_0=\frac{Q^2+m_\pi-\frac{t}{4}}{2\sqrt{s}}$. Finally, the element of integration over the internal momentum, considering only the 
imaginary part of the amplitude, reads:
\eqs{d^4k=d\tau d\phi\left.\frac{1}{8 \sqrt{s}|\vect q|}\right|_{k_0=k'_0,\cos \theta=\cos \theta_0},}
with $|\vect q|=\frac{1}{2x}\sqrt{\frac{4sx^2Q^2+(1-2x)^2Q^4}{s}}$. The boundary values of the 
integration domain on $\tau$ are obtained by solving $\cos \theta_0=\pm 1$.

Now we may look at the effect of the finite size 
of the pion on the integration procedure upon $k$.
The relative four-momentum squared of the quarks inside the pion is given by

\eqs{O_1^\pm&=&\left(2k-p\pm\frac{\Delta}{2}\right)^2\nn\\&=&
2\tau + 2 m_q^2 - m_{\pi}^2+\frac{t}{2}\pm 2k\cdot \Delta,\label{O1}}

for pion-quark vertices like the ones in diagram Fig.~\ref{fig:diagrams}.(a), and by

\eqs{&&O_2^\pm=\left(2k-p+2q\pm\frac{\Delta}{2}\right)^2\nn\\&=&-2\tau+6 m_q^2 - 
m_{\pi}^2+\frac{t}{2}-\frac{2Q^2}{x}\pm2(k\cdot \Delta+\frac{\xi Q^2}{x}),\label{O2}}

for vertices as in diagram  Fig.~\ref{fig:diagrams}.(b). 
Note that $k\cdot \Delta$ is a known function
of the external variables as well as of $\theta$ and $\tau$.
Generalising the procedure of \cite{ours}, we require either $|O_1^{\pm}| <
\Lambda^2$ or  $|O_2^{\pm}| < \Lambda^2$ for all diagrams. Gauge invariance is preserved
by the cut-off, since $|O_i^{\pm}|$ depend only upon the external variables of the $\gamma^\star\pi\to q \bar q $ process.

As the $O_i$'s and $\tau$ are always negative, we require one of the two following conditions:

\eqs{\tau &>& \frac{-\Lambda^2}{2}+ \frac{m_{\pi}^2}{2}-m_q^2-\frac{t}{4}+\left|k\cdot \Delta\right|,
\nn\\\tau &<& \frac{\Lambda^2}{2}- \frac{m_{\pi}^2}{2}+3m_q^2+\frac{t}{4}-\frac{Q^2}{x}
-\left|\frac{\xi Q^2}{x}+k\cdot \Delta\right|.
}

For $t$ small, $|O_1|$ and $|O_2|$ cannot be small simultaneously. 
The crossed diagrams
have their main contribution for $O_1\simeq O_2$, and are thus suppressed
by a power $\Lambda^2/Q^2$ when the cut-off is imposed. The box diagrams
have a leading contribution for $|O_1|$ or $|O_2|$ small, and are not
power suppressed by the cut-off.

It may be worth pointing out that the vertical propagators are more off-shell
in DVCS than in DIS, hence one would expect DVCS to be better described
by perturbation theory than DIS.

\subsection{The coupling constant}

In the diagonal case, we have determined the coupling constant $g=g_{\pi qq}$ by imposing 
that there are only two constituents in the pion. This sum rule constraints $F_1$ as follows~\cite{ours}:

\eqs{
\label{normF1}
\int _{0}^{1}F_{1}(x)dx=\frac{5}{18}.
}

As $F_1/g^2$ is a priori a function of $Q^2$, the sum rule imposes that 
$g$ should be a function of $Q^2$. 
But at high enough $Q^2$, where the details of the non-perturbative 
interaction are less and less relevant, $F_1/g^2$ reaches its asymptotic shape when the 
cut-off procedure is applied, and we obtain a constant value for $g$.
In Ref.~\cite{ours}, this asymptotic regime was reached for $Q^2$ as small as 2~GeV$^2$.
In the following, we shall make use of these previously obtained values, which 
are functions of the cut-off $\Lambda$.

However in the DVCS case, an ambiguity may arise as one of the vertices has an external 
kinematics similar to a vanishing $Q^2$ DIS. This ambiguity is lifted if one 
notices that the pertinent quantities are not $q_1^2$ and $q_2^2$ separately 
but the factorisation scale, which may be taken as 
the square of their mean, $Q^2$. Thus in DVCS, 
although $q_2^2$ vanishes, $Q^2$ does not and we shall consider that $g$ is constant.

\section{Results}
\subsection{General features}

\begin{figure*}
\centering
\mbox{
      \subfigure[]{\psfig{figure=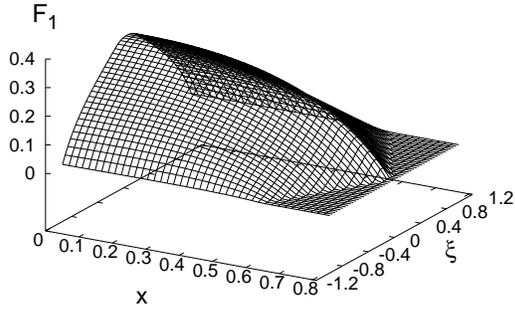,width=8.0cm,clip=true}}
      \subfigure[]{\psfig{figure=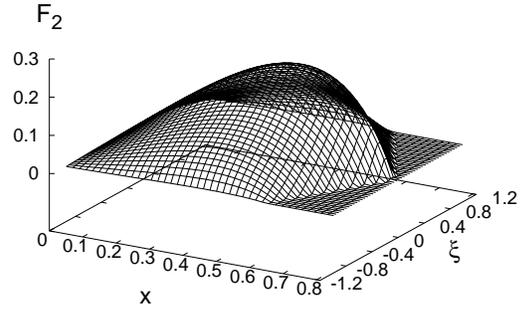,width=8.0cm,clip=true}}}
\mbox{\subfigure[]{\psfig{figure=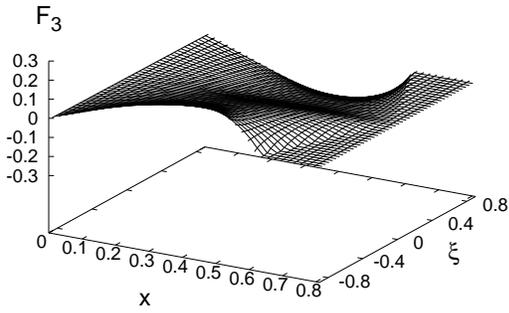,width=8.0cm,clip=true}}
      \subfigure[]{\psfig{figure=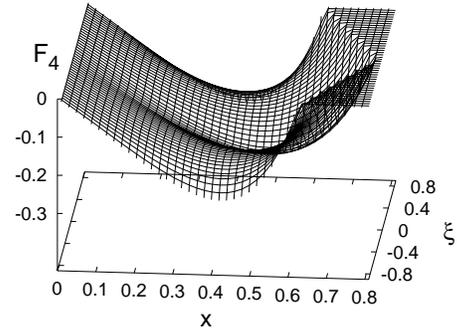,width=8.0cm,clip=true}}}
\mbox{\subfigure[]{\psfig{figure=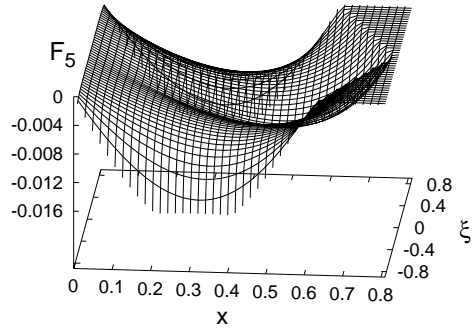,width=8.0cm,clip=true}}}
\caption{Plot of the five structure functions as functions of $x$ ($\in[0,1]$)and $\xi$ ($\in[-1,1]$) with a cut-off $\Lambda=0.75$ GeV, at $Q^2=10$~GeV$^2$
and $t=-0.1$ GeV$^2$.}
\label{fig:plot3d}
\end{figure*}

From the imaginary part of the total amplitude, the five structure functions $F_i$ 
can be obtained by a projection on the corresponding tensors. From now on, $F_i$ will stand
for the imaginary part of these structure functions.
In order to display their general features, we plot them 
in Fig.~\ref{fig:plot3d} first as functions of $x$ and $\xi$ for parameter 
values $m_q=0.3$ GeV and $\Lambda=0.75$~GeV, to ease the comparison with~\cite{ours}, 
and for $Q^2=10$~GeV$^2$ and $t=-0.1$~GeV$^2$. Let us notice that for any fixed value of $\xi$ not close 
to $\pm 1$, we recover for $F_1$ and $F_2$ the same behaviour as in the diagonal case. 
We checked indeed that the diagonal limit is recovered  for $\xi=0$ and $t=0$. 
Let us notice also that the structure functions 
$F_3$, $F_4$, $F_5$ depend little on $\xi$ except when this variable is close to $\pm 1$.

\begin{figure*}
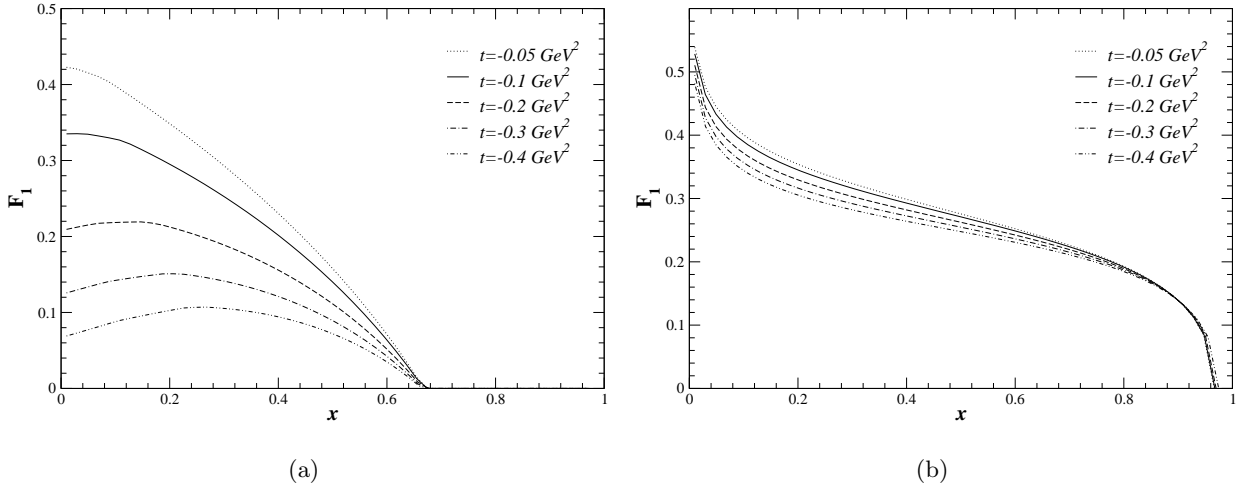

\centering
\mbox{
      \subfigure[]{\psfig{figure=F1dvcs-q2_10-lam_0_75.eps,width=8.0cm,clip=true}}\quad
      \subfigure[]{\psfig{figure=F1dvcs-nco-q2_10.eps,width=8.0cm,clip=true}}}

\caption{$F_1$ as a function of $x$ for various values of $t$ in the DVCS case with (a) and without (b) cut-off ($\Lambda=0.75$ GeV), at $Q^2=10$ GeV$^2$.}
\label{fig:F1_dvcs-tvar}
\end{figure*}

Let us turn now to DVCS. Fig.~\ref{fig:F1_dvcs-tvar} displays the behaviour of 
$F_1$ for various values of $t$ with and without cut-off. In the presence of 
size effects, the value of $F_1$ gets significantly 
reduced, especially for small $x$,  as $|t|$ increases, whereas that effect is 
much less noticeable  without cut-off.

\begin{figure*}
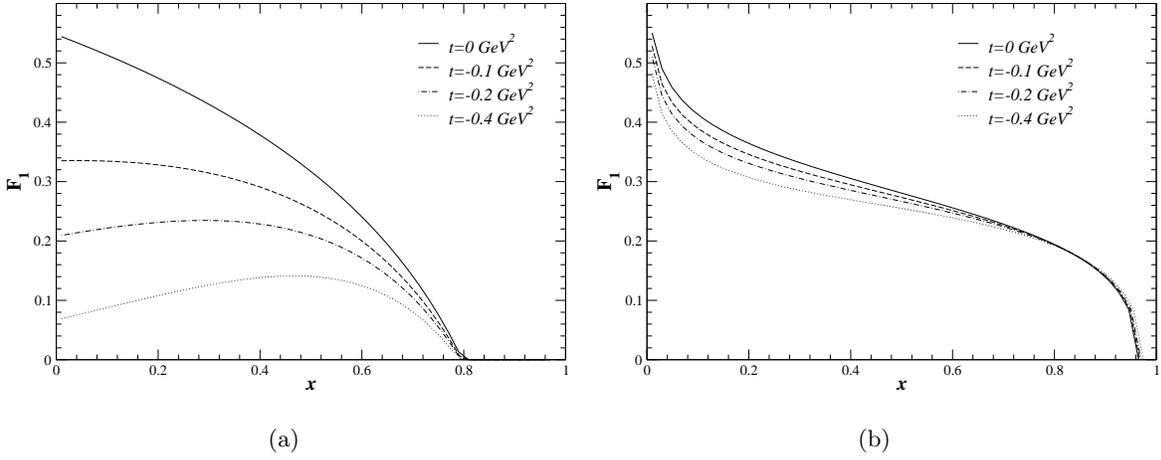

\centering

\mbox{
      \subfigure[]{\psfig{figure=F1-xi_0-q2_10-lam_0_75.eps,width=7.5cm,clip=true}}\quad
      \subfigure[]{\psfig{figure=F1-nco-xi_0-q2_10.eps,width=7.5cm,clip=true}}}
\caption{Evolution of $F_1$ (elastic case, $\xi=0$) for decreasing values of $t$ with (a) and without (b) cut-off, for $Q^2=10$ GeV$^2$.}
\label{fig:lim_diago}
\end{figure*}

\begin{figure*}
\centering

\psfig{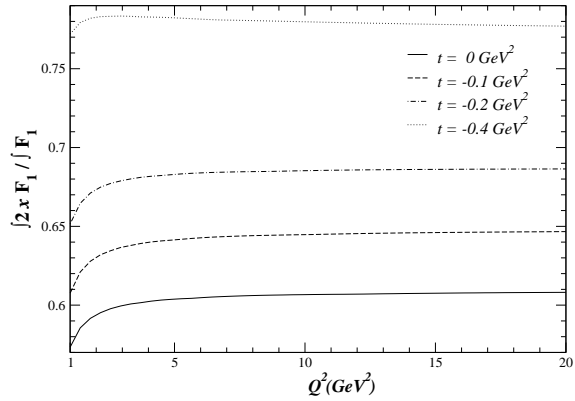}

\caption{Mean value of $2x$ for the $F_1$ distribution in the elastic case ($\xi=0$) as a function of $Q^2$ and for various values of $t$.}
\label{fig:intx}
\end{figure*}

In the elastic case (see Fig.~\ref{fig:lim_diago}), the same suppression at small $x$ is observed, especially
when the cut-off is applied. In Fig.~\ref{fig:intx}, we display the average value of $2x$ with respect to the $F_1$
distribution. The value of $\left<2x \right>$
increases when $|t|$ increases. The momentum fraction carried by the quarks and 
probed by the process thus increases with the momentum transfer. 

\begin{figure*}
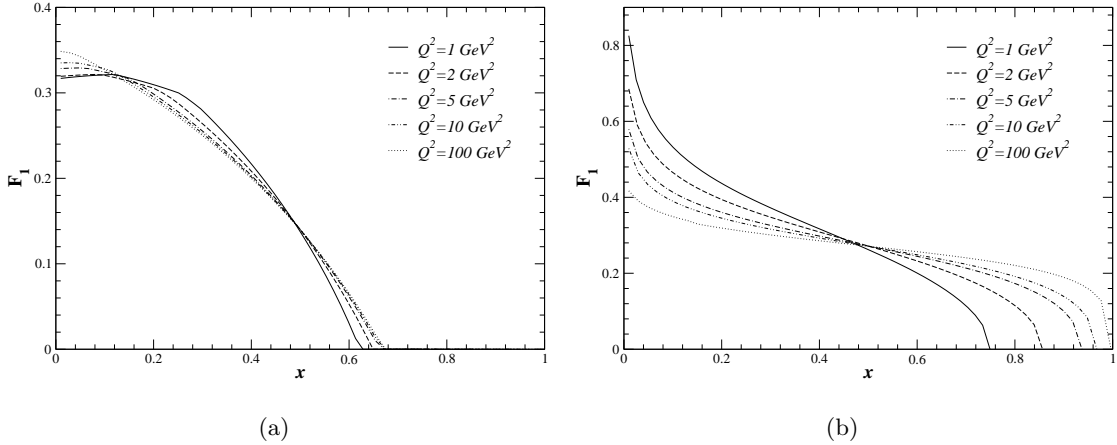

\centering

\mbox{
      \subfigure[]{\psfig{figure=F1dvcs-t_0_1-lam_0_75.eps,width=7.2cm,clip=true}}\quad
      \subfigure[]{\psfig{figure=F1dvcs-nco-t_0_1.eps,width=7.2cm,clip=true}}}
\caption{$F_1$ as a function of $x$ for $t=-0.1$ GeV$^2$ and for various values of $Q^2$ in the DVCS case with (a) and without (b) cut-off.}
\label{fig:F1_dvcs-q2var}
\end{figure*}

The effects of the variation of $Q^2$ are displayed in Fig.~\ref{fig:F1_dvcs-q2var}. As in the 
diagonal case~\cite{ours}, we can conclude that the details of the non-perturbative effects
cease to matter for $Q^2$ greater than 2 GeV$^2$, that is significantly larger than $\Lambda^2$. 
On the other hand, when the cut-off is not applied, we see (Fig.~\ref{fig:F1_dvcs-q2var} 
(b)) that $F_1$ evolves so slowly with $Q^2$ that the asymptotic state is not visible.

\subsection{High-$Q^2$ limit: new relations}

Having determined the 5 functions $F_i$'s in the context of our model, we 
shall now consider their behaviour at high $Q^2$. Expanding the ratios of 
$\frac{F_2}{F_1}$$,\frac{F_3}{F_1}$,$\frac{F_4}{F_1}$,$\frac{F_5}{F_1}$, 
we obtain the following asymptotic behaviour:
\eqs{
F_2&=&2xF_1+{\cal O}(1/Q^2), \label{eq:GCG1}    \\
F_3&=&\frac{2x\xi}{\xi^2-1}F_1+{\cal O}(1/Q^2),  \label{eq:GCG2}\\
F_4&=&\frac{2x}{\xi^2-1}F_1+{\cal O}(1/Q^2),  \label{eq:GCG3}         \\
F_5&=&{\cal O}(1/Q^2).\label{eq:GCG4} }

The fact that at leading order there are only three independent 
structure functions has 
been known 
for some time \cite{Belitsky:2000vk,GPD_pion1}. However, we show here that they can all 
be obtained from $F_1$.
The first relation is similar (at leading order in $1/Q^2$ and with the replacement of $x$ by $x_B$) to the Callan-Gross relation 
between the diagonal structure functions $F_1$ and $F_2$, valid for spin one-half 
constituents in general. Except for $F_5$, which is small at large $Q^2$, these relations 
show that $F_2$, $F_3$ and $F_4$ are simply related to $F_1$ at leading order. 
These relations clearly display and therefore confirm the 
symmetries of these functions. Combining Eqs.~(\ref{eq:GCG2}) and~(\ref{eq:GCG3}), 
we have, at leading order, 
\eqs{\label{eq:F3F4}F_3=\xi F_4,} 
which confirms that $F_3$ is an odd function of $\xi$, while $F_4$ is even.

The simple relations between the $F_i$'s (at leading order) constitute a remarkable
result of our model. Furthermore, we checked that the term ${\cal O}(1/Q^2)$
in Eq.~(\ref{eq:GCG1}) is numerically quite small, even for moderate $Q^2$. 
One may wonder whether these results are  typical of our model  or more general. 

\section{Linking the $F_i$'s to ${H}$, ${H}^3$, and $\tilde {H}^3$}

\begin{figure*}
{\centering \resizebox*{4.9cm}{3.8cm}{\includegraphics{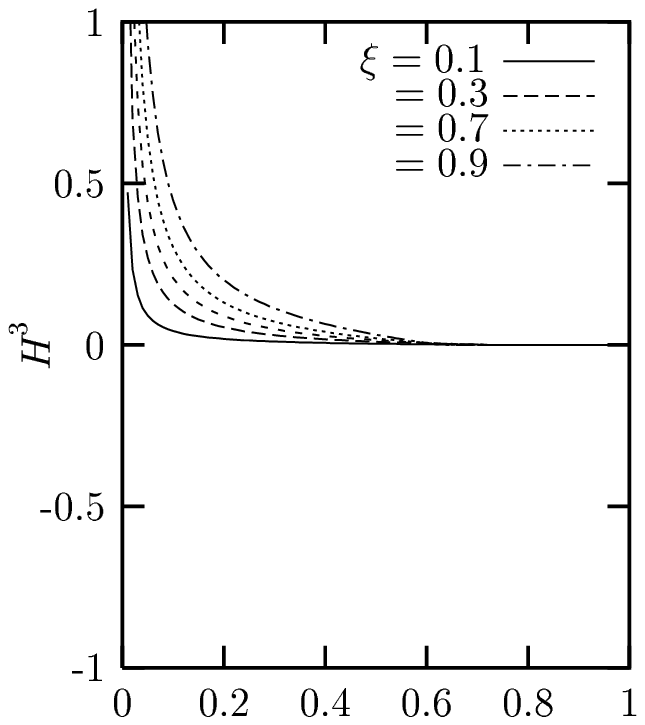}} \resizebox*{4.1cm}{3.8cm}{\includegraphics{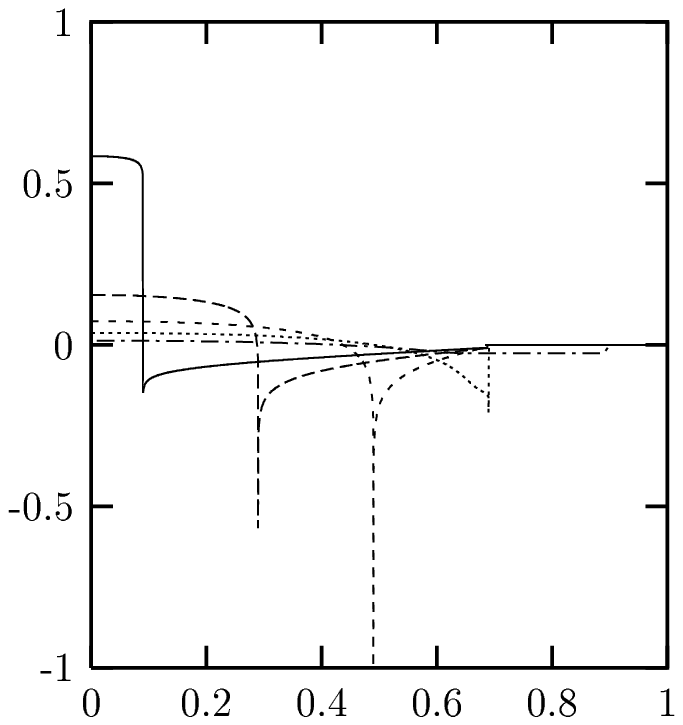}} \par}
\vspace{0.3cm}
~\vglue -1cm
\vspace{0.3cm}
\centering \resizebox*{4.9cm}{4.2cm}{\includegraphics{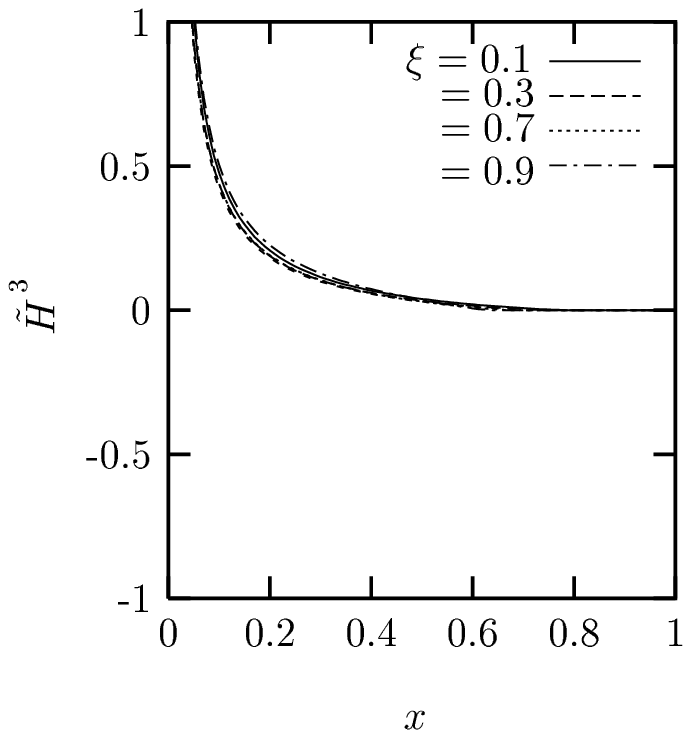}} \resizebox*{4.1cm}{4.2cm}{\includegraphics{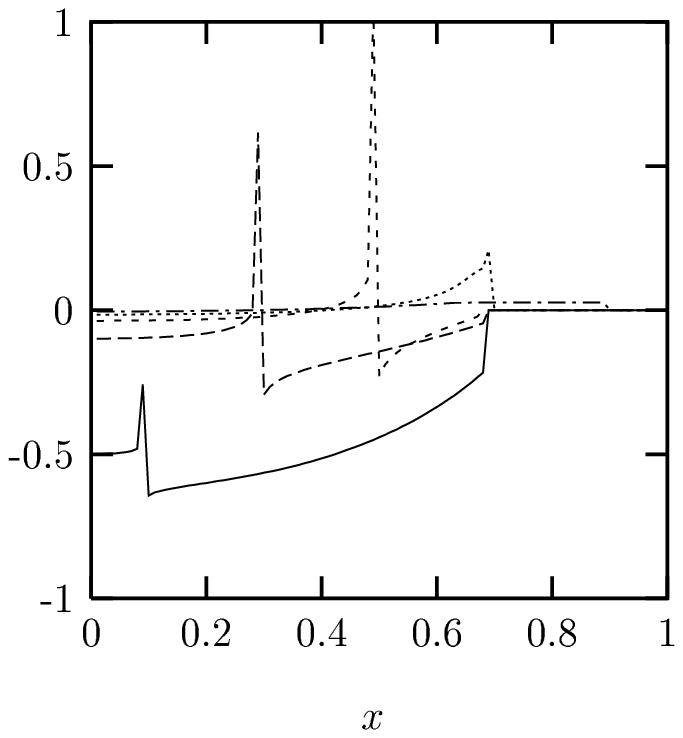}}
\caption{A comparison of the values of $H_3$ and $\tilde H_3$ obtained
in our model (left) with those calculated in the Wandzura-Wilczek
approximation, using
the value of $H$ from our model (right), for $Q^2=10$ GeV$^2$.}
\label{fig:WW}
\end{figure*}

Having at hand the five functions $F_i$'s that parametrise the amplitude for
$\gamma^\star\pi\to\gamma^\star\pi$, we would like to link them to the off-forward parton 
distribution functions or to the generalised parton distributions.
For this purpose, we make use of a tensorial expression coming from the twist-three analysis
 of the process, which singles out the twist-two ${\cal H}$ and the twist-three
 ${\cal H}^3,\tilde {\cal H}^3$ form factors. Following Ref.~\cite{Belitsky:2000vk}, we
write\footnote{Please note that Ref.~\cite{Belitsky:2000vk} uses ${\cal P}_{\nu\mu}$ instead of
${\cal P}_{\mu\nu}$ as projector.}:
\eqs{\label{Tw3Tamplitude}
T_{\mu\nu} (q, p, \Delta)=
&-& {\cal P}_{\sigma\mu} g^{\sigma\tau} {\cal P}_{\nu\tau}
\frac{q \cdot V_1}{2p \cdot q}\nn\\
&+& \left( {\cal P}_{\sigma\mu} p^\sigma  {\cal P}_{\nu\rho}
+ {\cal P}_{\rho\mu}  p^\sigma {\cal P}_{\nu\sigma} \right)
\frac{V_{2}^{\rho}}{p \cdot q}\nn\\
&-& {\cal P}_{\sigma\mu} i\epsilon^{\sigma \tau q \rho} {\cal P}_{\nu\tau}
\frac{A_{1\, \rho}}{2p \cdot q}.
}
where the $V_i$'s and $A_1$ read
\eqs{\label{V1V2A1}
V_{1 \, \rho} &=& 2 p_\rho {\cal H} +
(\Delta_\rho -2 \xi p_\rho)  {\cal H}^3 
+ \mbox{twist 4},\\
A_{1 \, \rho} &=& \frac{i\epsilon_{\rho\Delta p q}}{p\cdot q}\;
{\tilde{\cal H}}^3,\\
V_{2 \, \rho} &=& x V_{1 \, \rho} - \frac{x}{2}
\frac{p_\rho}{p\cdot q} q\cdot V_{1} + \frac{i}{4}
\frac{\epsilon_{\rho\sigma\Delta q}}{p\cdot q} A_{1}^{\sigma}
+ \mbox{twist 4}.\nonumber \\
}
In Ref.~\cite{Belitsky:2000vk}, gauge invariance of Eq.~(\ref{Tw3Tamplitude})  beyond 
the twist-three accuracy was in fact restored by hand, contrarily to 
the present calculation for which the amplitude is explicitly gauge invariant.

To relate the $F_i$'s to the $\cal H$'s, we project the amplitude~(\ref{Tw3Tamplitude})
onto the five projectors contained in Eq.~(\ref{tmunudecomp}) and identify the results
with the $F_i$'s. Note that, in the neutral pion case, the imaginary part of
the form factors $\cal H$ directly gives the GPD's $H$, $H^3$ and $\tilde H^3$ up to a factor $2\pi$. As we have
kept the off-shellnesses of the photons arbitrary, we in fact can relate the 
imaginary parts of $F_i$ to the GPD's for arbitrary $x$ and $\xi$:
\eqs{
{1\over 2\pi}F_1&=&{H}, \label{eq:FH1}   \\
{1\over 2\pi}F_2&=&2x{H}+{\cal O}(\frac{1}{Q^2}), \label{eq:FH2}  \\
{1\over 2\pi}F_3&=&\frac{2x}{x^2-\xi^2}\left({H}^3x^2+\tilde{H}^3\xi x-{H}\xi\right)+{\cal O}(\frac{1}{Q^2}), \label{eq:FH3}    \\ 
{1\over 2\pi}F_4&=&\frac{2x}{x^2-\xi^2}\left({H}^3\xi x+\tilde{H}^3x^2-{H}x\right)+{\cal O}(\frac{1}{Q^2}), \label{eq:FH4}\\
{1\over 2\pi}F_5&=& {\cal O}(\frac{1}{Q^2}).\label{eq:FH5}}

Replacing the $F_i$'s by the
expressions~(\ref{eq:GCG1}-\ref{eq:GCG4}), we can write 
\eqs{\tilde{H}^3=\frac{(x-1)}{x(\xi^2-1)}H+{\cal O}(\frac{1}{Q^2}),}
and
\eqs{{H}^3&=&\frac{(x-1)\xi}{x(\xi^2-1)}H+{\cal O}(\frac{1}{Q^2})=\xi\tilde{H}^3+{\cal O}(\frac{1}{Q^2}).}

As $F_1$ to $F_4$ can be written in term of only one of them, {\it e.g.} $F_1$, it
is not surprising that ${H}^3$ and ${\tilde H}^3$ are simply related
to ${H}$. Note that polynomiality of the Mellin moments of $H$, $H^3$ and $\tilde H^3$, together with Eqs.~(27) and (28), imply that
$H$ must be a polynomial $P_H$ multiplying $\xi^2-1$. The fact that, as can be seen from Fig. 7,
$\tilde H_3$ is almost independent of
$\xi$ shows that $P_H$ is very close to a constant.

To convince ourselves that relations (27) and (28) are new, we have compared 
them to the Wandzura-Wilczek approximation \cite{WW}, given in the pion case
in \cite{Belitsky:2000vk,Polyakov}. First of all, it is well-known that
these relations are discontinuous at $\xi=\pm x$, which is not the
case for (27) and (28). Furthermore, we show in Fig. 7 the results of
the Wandzura-Wilczek approximation compared with our results. We see that the two 
are numerically very different. Hence the relations (27) and (28), derived
in an explicitly gauge-invariant model, do not come from "kinematical" 
twist corrections, but emerge from the dynamics of the spectator quark 
propagator and from finite-size effects. 

\section{Discussion and conclusion}

We have extended our previous model for the pion to investigate  the off-diagonal structure 
functions for this particular case. 
The introduction of a cut-off allows the crossed diagrams to behave
as higher-twists and to relate the imaginary part of the forward amplitude with quark
GPD's.

We used the formalism of Ref.~\cite{Belitsky:2000vk} in order to decompose the amplitude
along the relevant Lorentz tensors, to define five structure functions $F_i$, and to relate 
the latter to the GPD's ${H}$, ${H}^3$ and $\tilde{H}^3$ introduced
in the twist analysis. We have found that our results in the forward case are 
qualitatively preserved when departing from the forward limit.

Our investigation yields new results. In particular, 
we singled out new relations, which link the $F_i$'s in a simple manner at leading
order in $1/Q^2$. More intriguing, we found that the twist-three structure functions are
simply related to ${H}$ by relations that differ from the Wandzura-Wilczek approximation.

Although these relations are derived in the context of our simple model, it is possible that they can be 
extended to a more general case.

\section*{Acknowledgements}
The authors wish to thank E. Ruiz-Arriola, P. Guichon and M.~V. Polyakov for their useful comments.
This work has been performed in the frame of the ESOP collaboration 
(European Union contract HPRN-CT-2000-00130).

%\appendix

% \section{}
% \label{}


\begin{thebibliography}{99}


% notes:

% \bibitem{label2}
% If there is a note, it should come last:
% \bibitem{label3} \note
% \end{subbibitems}


\bibitem{ours}
F.~Bissey, J.~R.~Cudell, J.~Cugnon, M.~Jaminon, J.~P.~Lansberg and P.~Stassart,
%``A model for the pion structure function,''
Phys.\ Lett.\ B {\bf 547} (2002) 210
[arXiv:hep-ph/0207107];
%%CITATION = HEP-PH 0207107;%%
%\cite{Lansberg:2002tu}
%\bibitem{Lansberg:2002tu}
J.~P.~Lansberg, F.~Bissey, J.~R.~Cudell, J.~Cugnon, M.~Jaminon and P.~Stassart,
%``Pion structure function and violation of the momentum sum rule,''
AIP Conf.\ Proc.\  {\bf 660} (2003) 339
[arXiv:hep-ph/0211450].
%%CITATION = HEP-PH 0211450;%%


\bibitem{diagonal}
%\bibitem{SH93}
T.~Shigetani, K.~Suzuki and H.~Toki,
Phys.\ Lett.\ B {\bf 308} (1993) 383
[arXiv:hep-ph/9402286];
%%CITATION = HEP-PH 9402286;%%
%
%\bibitem{DA95}
R.~M.~Davidson and E.~Ruiz Arriola,
Phys.\ Lett.\ B {\bf 348} (1995) 163;
%%CITATION = PHLTA,B348,163;%%
H.~Weigel, E.~Ruiz Arriola and L.~P.~Gamberg,
Nucl.\ Phys.\ B {\bf 560} (1999) 383
[arXiv:hep-ph/9905329];
%%CITATION = HEP-PH 9905329;%%
%\cite{RuizArriola:2002wr}
%\bibitem{RuizArriola:2002wr}
E.~Ruiz Arriola,
%``Pion structure at high and low energies in chiral quark models,''
Acta Phys.\ Polon.\ B {\bf 33} (2002) 4443
[arXiv:hep-ph/0210007];
%%CITATION = HEP-PH 0210007;%%
%\cite{Maris:2003vk}
%\bibitem{Maris:2003vk}
P.~Maris and C.~D.~Roberts,
%``Dyson-Schwinger equations: A tool for hadron physics,''
Int.\ J.\ Mod.\ Phys.\ E {\bf 12} (2003) 297
[arXiv:nucl-th/0301049];
%%CITATION = NUCL-TH 0301049;%%
%\cite{Detmold:2003tm}
%\bibitem{Detmold:2003tm}
W.~Detmold, W.~Melnitchouk and A.~W.~Thomas,
%``Parton distribution functions in the pion from lattice QCD,''
Phys.\ Rev.\ D {\bf 68} (2003) 034025
[arXiv:hep-lat/0303015].
%%CITATION = HEP-LAT 0303015;%%
\bibitem{experim}
%
%\bibitem{GRS}
M.~Gl\"uck, E.~Reya and I.~Schienbein,
Eur.\ Phys.\ J.\ C {\bf 10} (1999) 313
[arXiv:hep-ph/9903288].
%%CITATION = HEP-PH 9903288;%%
\bibitem{extramuller}
D.~M{\"u}ller, D.~Robaschik, B.~Geyer, F.~M.~Dittes and J.~Ho\v{r}ej{\v s}i,
 %``Wave functions, evolution equations and evolution kernels from light-ray
%operators of {QCD},''
Fortsch.\ Phys.\  {\bf 42} (1994) 101
[arXiv:hep-ph/9812448].
%%CITATION = HEP-PH 9812448;%%
%\cite{Belitsky:2000vk}
\bibitem{Belitsky:2000vk}
A.~V.~Belitsky, D.~M\"uller, A.~Kirchner and A.~Sch\"afer,
%``Twist-three analysis of photon electroproduction off pion,''
Phys.\ Rev.\ D {\bf 64} (2001) 116002
[arXiv:hep-ph/0011314].
%%CITATION = HEP-PH 0011314;%%


\bibitem{JI97}
X.~D.~Ji,
Phys.\ Rev.\ Lett.\  {\bf 78} (1997) 610
[arXiv:hep-ph/9603249];
%%CITATION = HEP-PH 9603249;%%
Phys.\ Rev.\ D {\bf 55} (1997) 7114
[arXiv:hep-ph/9609381].
%%CITATION = HEP-PH 9609381;%%
%
%
\bibitem{RA96}
A.~V.~Radyushkin,
Phys.\ Lett.\ B {\bf 380} (1996) 417
[arXiv:hep-ph/9604317];
%%CITATION = HEP-PH 9604317;%%
Phys.\ Rev.\ D {\bf 56} (1997) 5524
[arXiv:hep-ph/9704207].
%%CITATION = HEP-PH 9704207;%%

\bibitem{CO97}
J.~C.~Collins, L.~Frankfurt and M.~Strikman,
Phys.\ Rev.\ D {\bf 56} (1997) 2982
[arXiv:hep-ph/9611433].
%%CITATION = HEP-PH 9611433;%%

\bibitem{GPD_review}
%\cite{Guichon:1998xv}
%\bibitem{Guichon:1998xv}
P.~A.~Guichon and M.~Vanderhaeghen,
%``Virtual Compton scattering off the nucleon,''
Prog.\ Part.\ Nucl.\ Phys.\  {\bf 41} (1998) 125
[arXiv:hep-ph/9806305];
%%CITATION = HEP-PH 9806305;%%
%
%\cite{Goeke:2001tz}
%\bibitem{Goeke:2001tz}
K.~Goeke, M.~V.~Polyakov and M.~Vanderhaeghen,
%``Hard exclusive reactions and the structure of hadrons,''
Prog.\ Part.\ Nucl.\ Phys.\  {\bf 47} (2001) 401
[arXiv:hep-ph/0106012];
%%CITATION = HEP-PH 0106012;%%
%
%\cite{Diehl:2003ny}
%\bibitem{Diehl:2003ny}
M.~Diehl,
%``Generalized parton distributions,''
Phys.\ Rept.\ {\bf 388} (2003) 41
[arXiv:hep-ph/0307382].
%%CITATION = HEP-PH 0307382;%%
\bibitem{GPD_pion1}
A.~V.~Radyushkin and C.~Weiss,
Phys.\ Rev.\ D {\bf 63} (2001) 114012
[arXiv:hep-ph/0010296];
%%CITATION = HEP-PH 0010296;%%
Phys.\ Lett.\ B {\bf 493} (2000) 332
[arXiv:hep-ph/0008214].
%%CITATION = HEP-PH 0008214;%%
\bibitem{GPD_pion}
%\cite{Broniowski:2003rp}
%\bibitem{Broniowski:2003rp}
W.~Broniowski and E.~Ruiz Arriola,
%``Impact-parameter dependence of the generalized parton distribution of the pion in chiral quark models,''
arXiv:hep-ph/0307198;
%%CITATION = HEP-PH 0307198;%%
%
%\cite{Dalley:2003sz}
%\bibitem{Dalley:2003sz}
S.~Dalley,
%``Impact parameter dependent quark distribution of the pion,''
Phys.\ Lett.\ B {\bf 570} (2003) 191
[arXiv:hep-ph/0306121];
%%CITATION = HEP-PH 0306121;%%
%
%\cite{Theussl:2002xp}
%\bibitem{Theussl:2002xp}
L.~Theussl, S.~Noguera and V.~Vento,
%``Generalized parton distributions of the pion in a Bethe-Salpeter  approach,''
arXiv:nucl-th/0211036;
%%CITATION = NUCL-TH 0211036;%%
A.~E.~Dorokhov and L.~Tomio,
%``Pion Structure Function Within The Instanton Model,''
Phys.\ Rev.\ D {\bf 62} (2000) 014016 [arXiv:hep-ph/9803329];
%%CITATION = PHRVA,D62,014016;%%
I.~V.~Anikin, A.~E.~Dorokhov, A.~E.~Maksimov, L.~Tomio and V.~Vento,
%``Nonforward Parton Distributions Of The Pion Within An Effective Single
%Instanton Approximation,''
Nucl.\ Phys.\ A {\bf 678} (2000) 175 [arXiv:hep-ph/9905332];
%%CITATION = NUPHA,A678,175;%%
\bibitem{WW}
S.~Wandzura and F.~Wilczek,
Phys.\ Lett.\ B {\bf 72} (1977) 195.
%%CITATION = PHLTA,B72,195;%%
\bibitem{Polyakov}
N.~Kivel, M.~V.~Polyakov, A.~Sch\"afer and O.~V.~Teryaev,
Phys.\ Lett.\ B {\bf 497} (2001) 73
[arXiv:hep-ph/0007315]. Note that the definitions of $H_3$ and $\tilde H_3$
differ by a factor $-2$ from those of \cite{Belitsky:2000vk} which we follow.
%%CITATION = HEP-PH 0007315;%%
\end{thebibliography}
\end{document}